\documentclass[fleqn,usenatbib,usedcolumn]{mnras}
  \usepackage{amsmath}
  \usepackage{upgreek}
   \usepackage{hyperref}
   \usepackage[british]{babel}             
   \usepackage{txfonts}                    
   \let\la=\lesssim 
   \usepackage{graphicx}                   

   \hypersetup{pdfauthor={Jos\'e Fonseca, Roy Maartens, M\'ario G. Santos},
               pdftitle={Probing the primordial Universe with MeerKAT and DES},
               pdfkeywords={cosmology: large-scale structure of the Universe, miscellaneous},
               bookmarksnumbered=true}
   \hypersetup{colorlinks=true,
            linkcolor=blue,
            citecolor=blue,
            filecolor=blue,
            urlcolor=blue}

\newcommand{\be}{\begin{equation}}
\newcommand{\ee}{\end{equation}}
\newcommand{\bea}{\begin{eqnarray}}
\newcommand{\eea}{\end{eqnarray}}
\newcommand{\bfig}{\begin{figure}}
\newcommand{\efig}{\end{figure}}
\newcommand{\nn}{\nonumber}

\def\fnl{{f_{\rm NL}}}
\def\d{\mathrm{d}}
\def\G{{\rm L}}
\def\l{\left(}
\def\r{\right)}
\def\D{\Delta}
\def\H{{\mathcal H}}
\def\p{\partial}
\def\fsky{f_{\rm sky}}
\def\z{(z)}
\def\s{{\sigma}}
\def\odm{\Omega_{\rm cdm}}
\def\ob{\Omega_{\rm b}}
\def\cs{{\sc camb}$\_$sources}

\title[The primordial Universe with MeerKAT and DES]{Probing the primordial Universe with MeerKAT and DES}
\author[Fonseca \textit{et al.}]{Jos\'e Fonseca$^1$\thanks{josecarlos.s.fonseca@gmail.com}, Roy Maartens$^{1,2}$ \& M\'ario G. Santos$^{1,3}$ \\
$^1$ Department of Physics \& Astronomy, University of the Western Cape, Cape Town 7535, South Africa\\
$^2$ Institute of Cosmology \& Gravitation, University of Portsmouth, Portsmouth PO1 3FX, UK\\
$^3$ SKA SA, The Park, Park Road, Cape Town 7405, South Africa\\
}

\date{}

\pagerange{\pageref{firstpage}--\pageref{lastpage}}

\begin{document}

\label{firstpage}

\maketitle

\begin{abstract}
It is usually assumed that we will need to wait until next-generation surveys like Euclid, LSST and SKA,  in order to improve on the current best constraints on primordial non-Gaussianity from the Planck experiment. We show that two contemporary surveys, with the SKA precursor MeerKAT and the Dark Energy Survey (DES), can be combined using the multi-tracer technique to deliver an accuracy on measurement of $\fnl$ that is up to three times better than Planck.

\end{abstract}



\section{Introduction}

Future cosmological surveys will probe the 3-dimensional large-scale structure of the Universe in ever larger volumes, delivering tighter and tighter constraints on cosmological parameters and modified gravity. Most of these surveys are based on sampling a large number of galaxies at optical or near-infrared wavelengths, such as Euclid\footnote{www.euclid-ec.org} and LSST\footnote{www.lsst.org}.
The SKA\footnote{www.skatelescope.org} will use the 21cm emission of HI, both to detect HI galaxies and to map the integrated intensity from each pixel.

One of the key targets of these next-generation surveys is to go beyond the capability of CMB experiments in probing the primordial Universe -- in particular to surpass CMB constraints on primordial non-Gaussianity. Primordial Non-Guassianity in the fluctuations that are generated by inflation leaves a `frozen' signal in the matter distribution on ultra-large scales, which is why ultra-large volume surveys will be able to improve on the CMB state of the art. 
In this paper we only focus on local-type non-Gaussianity, measured by the parameter $\fnl$. 
The current best bound on $\fnl$ is from the Planck experiment, giving\footnote{We use the large-scale structure convention, $\fnl=f_{\rm NL}^{\rm LSS} \simeq 1.3 f_{\rm NL}^{\rm CMB}$.} $\sigma(\fnl)\simeq 6.5$ \citep{Ade:2015ava}.

Recent and current surveys, such as BOSS\footnote{www.sdss3.org/surveys/boss.php} and DES\footnote{www.darkenergysurvey.org}, are unable to match the CMB accuracy on $\fnl$. Next-generation surveys are forecast to beat the Planck constraint  \citep{Alonso:2015uua,Raccanelli:2015vla}. Even these surveys, using the galaxy power spectrum and a single tracer,  will be unable to push $\sigma(\fnl)$ below 1,  which is needed to rule out some of the simplest inflationary models. The problem is cosmic variance, which grows with the increase of scales being probed.
A way to beat down cosmic variance is the multi-tracer technique, which combines the information from two or more surveys (or multiple tracers within the same survey) \citep{Seljak:2008xr,McDonald:2008sh,Hamaus:2011dq,Abramo:2013awa}. This technique is forecast to deliver game-changing improvements in $\sigma(\fnl)$ from next-generation surveys \citep{Yoo:2012se,Ferramacho:2014pua, Yamauchi:2014ioa,Alonso:2015sfa,2015ApJ...812L..22F}. 

However these improvements will only be available from future surveys in the coming decade. Hence a question arises: can surveys be combined within the next few years to match or improve CMB bounds on $\fnl$? 
We address this question, using the multi-tracer technique for two surveys. The multi-tracer confines us to use the overlap sky area and redshift range. Furthermore, the technique is more powerful the more different are the tracers of dark matter in each survey, and the more different are the systematics. This leads us to choose the two contemporary surveys: DES (optical/infrared telescope, photometry) and MeerKAT\footnote{www.ska.ac.za/science-engineering/meerkat/} (radio dish array, HI intensity mapping). Our forecast is that DES and MeerKAT combined can measure $\fnl$ with  Planck-level accuracy or better. 

In Section \ref{sec:fisher} we review the large-scale effects of non-Gaussianity as well as the Fisher forecast method using the multi-tracer technique. Then we describe in Section \ref{sec:surveys} the experimental specifications for MeerKAT and DES. In Section \ref{sec:results} we present our results and we conclude in Section \ref{sec:conc}.

\section{Theoretical  ingredients} \label{sec:fisher}

\subsection{Local primordial non-Gaussianity}

Local primordial non-Gaussianity is described by a nonlinear correction to the primordial Newtonian potential $\Phi(\bf x)$:
\bea
\Phi=\phi+\fnl (\phi^2-\langle \phi^2\rangle).
\eea
Here $\phi$ is a first-order Gaussian potential and the perturbed metric is ${\d}s^2=a^2[-(1+2\psi){\d}\eta^2 + (1-2\phi){\d}\boldsymbol{x}^2]$. The galaxy power spectrum is altered by scale-dependent bias on large scales \citep{Dalal:2007cu,Matarrese:2008nc}:
\bea\label{bng}
 b(z,k)=b_\G(z)+3\fnl \frac{\big[ b_{\G}(z) -1\big]\Omega_m H_0^2 \delta_c}{D(z) T(k)  k^2},
\eea
where $b_{\G}$ is the linear Gaussian bias, $\delta_c\simeq 1.69$ is the critical matter density contrast for spherical collapse, $T$ is the transfer function (normalised to 1 on large scales) and $D$ is the growth factor (normalised to 1 at $z=0$). On super-equality scales,  $T(k)\simeq 1$ and  the bias grows as $\fnl k^{-2}$. If we use the power spectrum to probe primordial non-Gaussianity we therefore need to look at the largest scales possible. When using a single tracer of the dark matter distribution, the signal is eroded by cosmic variance, and even the next-generation ultra-large survey volumes are unable to achieve $\sigma(\fnl)<1$ \citep{Alonso:2015uua,Raccanelli:2015vla}. We deal with the cosmic variance problem by using multiple tracers, following the method of \citet{2015ApJ...812L..22F}.

\subsection{Angular power spectrum with all relativistic effects}

The observed number density or brightness temperature contrast is $\Delta^A(z,\boldsymbol n)$, where $\boldsymbol n$ is the direction of observation and $A$ labels the tracer. Its two-point correlators define the angular power spectra: 
\bea
\big\langle \Delta^A(z,\boldsymbol n)\Delta^B(z',\boldsymbol n')\big\rangle =\sum_\ell {(2\ell+1)\over 4\pi}\, C_\ell^{AB}(z,z')\, P_\ell(\boldsymbol n \cdot \boldsymbol n'),
\eea
{where $P_\ell$ are the Legendre polynomials.} The sky maps of the tracers are decomposed into spherical harmonic modes and the $a_{\ell m}$ are used as estimators. We assume that the $a_{\ell m}$ are Gaussian distributed. Since the universe has no preferred structure ($\langle a_{\ell m} \rangle = 0$), all the information will be encoded in the angular power spectra $C^{AB}_\ell$, where $\langle a^A_{\ell m} a^{B*}_{\ell' m'}\rangle = \delta_{\ell \ell'}\delta_{m m'} C^{AB}_\ell$.

Extending the single-tracer case \citep{Challinor:2011bk} to multiple tracers, the angular power spectra across two redshift bins are given by
\bea \label{cl}
C^{AB}_\ell \!\l z_i,z_j\r=4\pi\!\!\int\!\!\d \ln k\,\D_\ell^{W_A}\!\l z_i,k\r \D_\ell^{W_B}\!\l z_j,k\r \mathcal P\!\l k\r \!,
\eea
where $z_i$ are the redshift bin centres and the dimensionless primordial curvature perturbation power spectrum is 
\bea
\mathcal P (k)=A_s\l \frac k{k_0}\r^{n_s-1}.
\eea
Here the pivot scale is $k_0=0.05\,$Mpc$^{-1}$,  $A_s$ is the amplitude  and $n_s$ is the spectral index. The theoretical transfer function $\D^{ A}_\ell(z,k)$ (not to be confused with $T(k)$) defines the
observable transfer function  in each bin via the window function $W$:
\bea \label{transf}
\D_\ell^{W_A}\l z_i,k\r=\int\!\! \d z\,p^A(z)W(z_i,z)\D_\ell^A(z,k).
\eea
Here $p^A$ is a selection function for tracer $A$ which is {de facto} the redshift distribution function of observed sources. For galaxy number counts, $p^A\propto \d n^A/\d z \d\Omega$. For HI temperature intensity maps, $p^A\propto T^A$.  
The selection function accounts for the fact that we have different numbers of emitters at different redshifts. It therefore weights the relative importance of each redshift in the signal. 

The observational window function centered on $z_i$ is $W(z_i,z)$, and is the probability distribution function for a source to be inside the $i$-bin. This is broadly speaking a binning choice based on the experimental specifications.  The window function can also be chosen differently for different tracers, but when using the multi-tracer technique it has to be the same. 
The product of the selection function and the window function is the tracer's effective redshift distribution function inside the bin, normalised so that $\int\!\d z\,p^A(z)W(z_i,z)=1$ for all $z_i$.

The observed fluctuations $\D^A(z,{\boldsymbol n})$, and thus the transfer functions $\D^{ A}_\ell(z,k)$, are gauge-independent and any gauge may be used to compute them. For galaxies, expressions have been given in \citet{Yoo:2010ni,Challinor:2011bk,Bonvin:2011bg} and for HI intensity mapping in \citet{Hall:2012wd}. In Newtonian gauge, we have \citep{DiDio:2013bqa,2015ApJ...812L..22F}
\bea
\D^{A}_\ell(k)&=&\left[ b^A\delta^{\rm s}_k+\l b^A_e-3\r\frac {\H v_k}{k}\right] j_\ell\l k\chi\r+\frac {k v_k}{\H}j_\ell''(k\chi)\nn\\
&&{}+\frac{\ell\l\ell+1\r\l2-5 s^A\r}2\int_0^{\chi}\d\tilde\chi\frac{(\chi-\tilde\chi)}{\chi\tilde\chi}\l\tilde\phi_k+\tilde\psi_k\r j_\ell\l k\tilde\chi\r\nn\\
&&{}+\l\frac{2-5s^A}{\H\chi} +5s^A-b^A_e+\frac{\H'}{\H^2}\r
  \bigg[ v_k j_\ell'(k\chi) +\psi_k j_\ell\l k\chi\r \nn\\&& + \int_0^{\chi}\d\tilde\chi\l\tilde\phi'_k+\tilde\psi'_k\r j_\ell\l k\tilde\chi\r\bigg]
\nn\\
&&{}+\frac{\l2-5s^A\r}{\chi}\int_0^{\chi}\d\tilde\chi\l\tilde\phi_k+\tilde\psi_k\r j_\ell\l k\tilde\chi\r\nn\\
&&{}+\left[ \psi_k+\l5s^A-2\r \phi_k+\frac{\phi_k'}{\H}\right]  j_\ell\l k\chi\r, \label{eq:angGS}
\eea
where we have suppressed the redshift dependence, $\cal H$ is the conformal Hubble parameter, $\chi$ is the comoving line-of-sight distance, and $v_k$ is the peculiar velocity. For $\Lambda$CDM and standard dark energy models, the metric perturbations are equal: $\psi_k=\phi_k$.

The first term on the right of \eqref{eq:angGS} is the contribution from the tracer fluctuations, where $\delta^{\rm s}$ is the dark matter density contrast in the matter restframe, equivalently in synchronous gauge. It is necessary to use the restframe in order to avoid gauge-dependence in the definition of bias \citep{Challinor:2011bk,Bonvin:2011bg,Bruni:2011ta,Jeong:2011as}. In the presence of non-Gaussianity, the bias $b^A(z,k)$ of tracer $A$ is given by \eqref{bng}, with $b_\G\to b_\G^A$.  The evolution bias, $b^A_e(z)$ accounts for the redshift evolution of the sources for tracer $A$:
\bea \label{be}
b^A_e= -\frac{\p\ln \big[(1+z)^{-3}N^A\big]}{\p\ln (1+z)}\,,
\eea
where $N^A$ is the background number density, of galaxies or HI atoms. 

The second term on the right of \eqref{eq:angGS} is the redshift-space distortion contribution,  which is independent of the chosen tracer (given the assumption that there is no velocity bias). 

The second line of \eqref{eq:angGS} gives the  contribution of lensing convergence to the tracer fluctuations, integrated along the line of sight to each source. The lensing effect is modified by the magnification bias, $s^A$.
Here we need to make a careful distinction between number counts and intensity mapping. At linear order, there is no lensing contribution to the HI intensity fluctuations. This follows from conservation of surface brightness in gravitational lensing, similar to the case of CMB temperature fluctuations.
It turns out that the HI brightness temperature fluctuations coincide with the HI atom number density fluctuations, provided that we set $s^{HI}=2/5$, which removes the lensing contribution, and some other terms in \eqref{eq:angGS} related to the luminosity distance fluctuations \citep{Hall:2012wd}. For galaxy number counts, $s^G$ is the logarithmic slope of the cumulative luminosity function $N^A(z,m<m_\ast)$ at the magnitude limit $m_\ast$ of the survey. Thus we have
\bea
s^{HI} &=& {2\over5}, \label{sh}\\
s^G&=&\frac{\p \log_{10} N^G} {\p m_\ast}. \label{sg}
\eea

In the third line of \eqref{eq:angGS}, there is a Doppler term and a Sachs-Wolfe term. The fourth and fifth lines are integrated Sachs-Wolfe and time delay terms respectively,  while the final line is a further Sachs-Wolfe type contribution. 
The last 4 lines of \eqref{eq:angGS} are the horizon-scale relativistic effects, which have $k$-dependences of $(\H/k)\,\delta^{\rm s}_k$ and $(\H/k)^2\delta^{\rm s}_k$. This follows from the Euler equation, which gives $v_k=f(\H/k)\,\delta^{\rm s}_k$, and the Poisson equation, which gives $\phi_k\propto (\H/k)^2\delta^{\rm s}_k$.
Note that the first line also contains an horizon-scale term, since ${\cal H}v_k/k=f(\H/k)^2\delta^{\rm s}_k$. 

These relativistic terms become relevant on the same ultra-large scales where local primordial non-Gaussianity is boosting the power spectrum via the bias, \eqref{bng}. For accurate constraints on $\fnl$, we need to include the relativistic terms. In the single-tracer case, this has been done by \cite{Camera:2014bwa,Alonso:2015uua,Raccanelli:2015vla}. Note that the best-fit value of $\fnl$, as opposed to the measurement error $\sigma(\fnl)$, can be significantly biased if the relativistic terms are omitted \citep{Camera:2014sba}.  For the multi-tracer case, the relativistic effects can be detected and simultaneously $\sigma(\fnl)<1$ can be achieved, as shown by  \cite{Yoo:2012se} when neglecting all integrated terms in \eqref{eq:angGS}, and by \cite{Alonso:2015sfa,2015ApJ...812L..22F} in the general case.

\subsection{Fisher forecasts with multiple tracers}

The Fisher matrix for a set of parameters $\{\vartheta_i\}$ is
\bea \label{eq:fishergeneral} 
{ F}_{\vartheta_i\vartheta_j}=\frac12 {\rm Tr}\left[ \left(\partial_{\vartheta_i}{ C}\right) {\Gamma}^{-1}\left(\partial_{\vartheta_j} {  C}\right){ \Gamma}^{-1}\right], ~~~{ \Gamma}={  C}+{  \cal N},
\eea
where ${C}$ is the covariance matrix of the estimator and ${ \cal N}$ is the noise contaminant, which we assume is independent of $\fnl$. For instrumental noise in radio surveys this is necessarily true, but it may not hold for shot noise, since non-Gaussianity induces deviations in halo over-density from the pure Poisson sampling noise case -- see \cite{Hamaus:2011dq} for a discussion. However the same authors conclude that this correction is tiny if one considers a large halo-mass bin, which will be the case in this paper, hence justifying our assumption. If the angular power spectrum $C^{AB}_\ell(z_i,z_j)$ is the estimator's covariance,  then we need to account for all multipoles and \eqref{eq:fishergeneral} becomes \citep{Tegmark:1996bz}
\bea \label{eq:fishercl}
{ F}_{\vartheta_i\vartheta_j}=\sum_{\ell_{\rm min}}^{\ell_{\rm max}} \frac{(2\ell+1)}{2 }\fsky\,{\rm Tr}\left[ \left(\partial_{\vartheta_i}{  C}_{\ell}\right){ \Gamma}_\ell^{-1} \left(\partial_{\vartheta_j}{  C}_{\ell}\right){ \Gamma}_\ell^{-1}\right]\,,
\eea
where $\fsky$ is the fraction of sky surveyed. The multi-tracer technique requires that we use the same sky maps of two (or more) differently biased tracers. Not only should the sky areas be the same, but also the binning in redshift, so that we are always comparing the same patch of the universe. If we use an HI intensity map survey and a galaxy survey, we can schematically represent the covariance matrix as \citep{Ferramacho:2014pua} 
\bea
{ C}^{AB}_\ell\l z_i,z_j\r
=
\begin{pmatrix}
    {  C}^{HI,HI}_{\ell,ij} & { C}^{HI,G}_{\ell,ij} \\ &&\\
   { C}^{G,HI}_{\ell,ij} &  { C}^{G,G}_{\ell, ij} 
\end{pmatrix}\,.
\eea
Note that if we have $n$ bins then the covariance matrix is $2n\times 2n$ for two tracers. The auto-tracer correlations are symmetric, but not the cross-correlations. Nevertheless the overall angular power is symmetric with ${ C}^{ HI,G}_{\ell,ij}=C^{G,HI}_{\ell,ji}$. We do not include foregrounds and systematics in the full covariance matrix $\Gamma$. Note that, in addition to the reduction of cosmic variance, the multi-tracer technique also lessens the individual systematics of the two experiments and reduces the impact of foreground residuals.  

Assuming that for Gaussian likelihoods the inverse of the Fisher matrix approximates well the parameter covariance,  the marginal error in a parameter is  given by
\bea
\sigma_{\vartheta_i}=\big[{({  F}^{-1})_{\vartheta_i\vartheta_i}}\big]^{1/2}\,.
\eea

\section{Surveys}\label{sec:surveys}

The multi-tracer technique is more effective if the differences between the tracers, and between the experimental characteristics, are large. An intensity map in the radio and a photometric galaxy survey have very different experimental features and the bias, evolution bias and magnification bias are also very different. We combine the two premier contemporary surveys of these types --
an HI intensity survey with MeerKAT and a galaxy survey with DES.

\subsection{MeerKAT HI intensity map}

MeerKAT will be composed of 64 antennas and will operate from 2017. A proposed cosmological survey MeerKLASS \citep{Santos:2016xxx} includes an HI intensity map survey. Forecasts for such a survey have been investigated  
\citep{Bull:2015lja, Pourtsidou:2015mia, Pourtsidou:2016dzn}, showing that MeerKAT can provide very good cosmological  constraints. 
 
In HI intensity mapping, all galaxies with HI contribute to the signal. To compute the angular power spectrum we use the  transfer function multipoles given by \eqref{eq:angGS} together with \eqref{sh}. For brightness temperature fluctuation maps, the selection function follows the HI temperature, $p^{HI}(z)\propto T^{ HI}(z)$, which we fit using the results of \citet{Santos:2015bsa}. 

\begin{figure}
\centering
\includegraphics[width=\columnwidth]{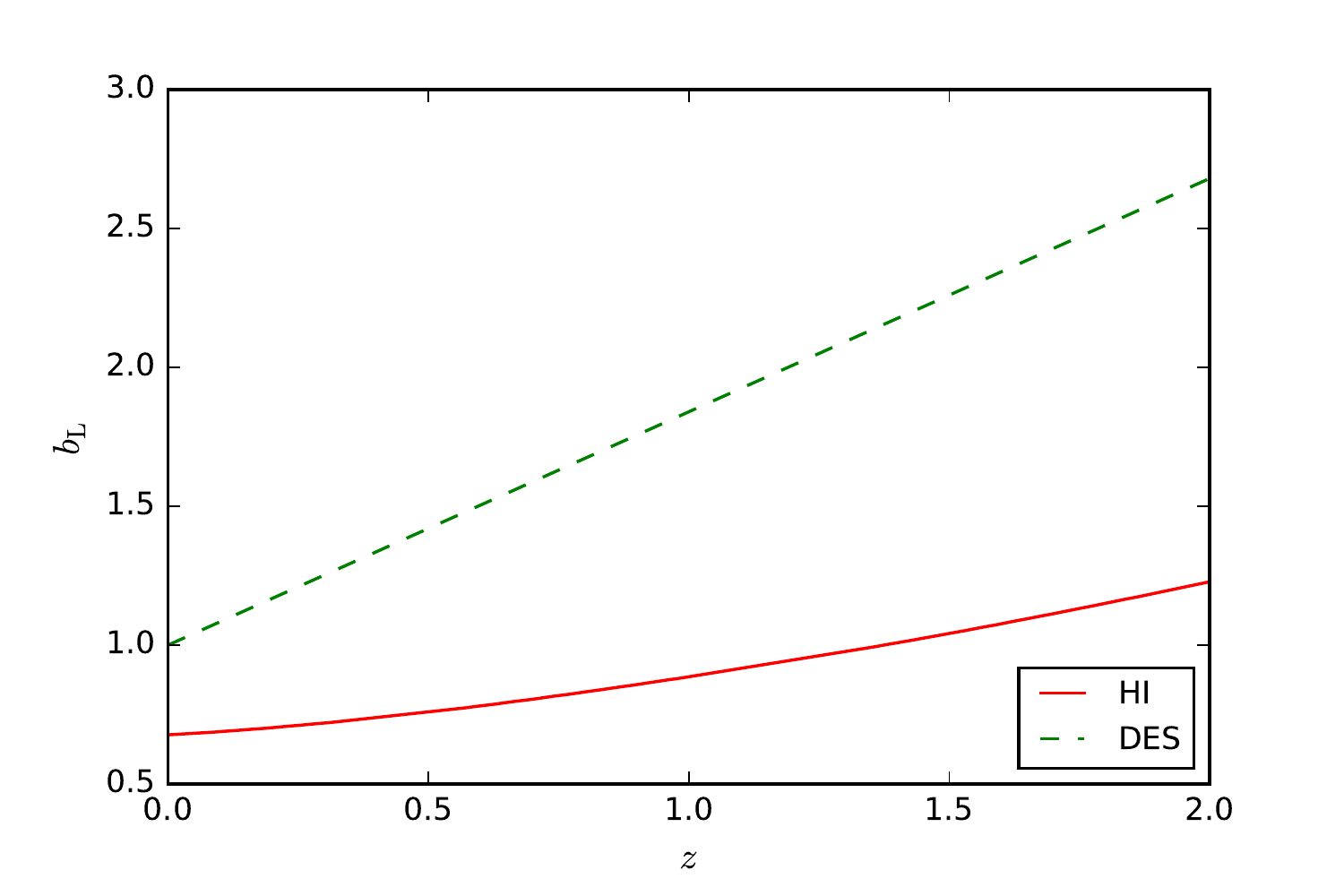}
\caption{Gaussian bias for HI intensity (solid, red) and DES (dashed, green).}\label{fig:bias}
\end{figure}

The Gaussian HI bias, $b_{\rm L}^{HI}$, is modelled by weighting the halo bias with the HI content in the dark matter haloes, and is shown in Fig. \ref{fig:bias}: details of the modelling are given in \cite{Santos:2015bsa}. In our forecasts, we marginalize over the HI and galaxy bias -- we use the modelled bias only to set the fiducial bias value in each redshift bin.

Given the background relation between $N^{HI}$ and $T^{ HI}$, the HI evolution bias  \eqref{be} becomes \citep{Hall:2012wd}
\bea
b_e^{ HI}(z)=-\frac{\p\big[\ln T^{ HI}\z \H\z (1+z)^{-1}\big]}{\p\ln (1+z)}\,.
\eea

The noise angular power spectrum for an experiment with $N_{\rm d}$ collecting dishes, total observation time $t_{\rm tot}$ and observed sky area $S_{\rm area}$, is given by
\bea
\mathcal N^{ HI}_{ij}=\frac{S_{\rm area}}{2N_{\rm d} t_{\rm tot}}\int\!\d \nu\, T^2_{\rm sys}\l\nu\r \bar W_\nu\l\nu,\nu_i\r \bar W_\nu( \nu,\nu_j) \,,\label{eq:N_HI}
\eea
where $T_{\rm sys}$ is the system temperature of the receiver. The window function in frequency is equivalent to the one in redshift, given that $\bar W_\nu(\nu,\nu_i)=\bar W\l z,z_i\r \d z/\d\nu$ and that the window is normalised: $\int\!\d \nu\,\bar W_\nu(\nu,\nu_i)=1$. The expression \eqref{eq:N_HI} is more general than what is commonly found in the literature and is valid when we do not consider a top hat window function. This allows us to deal correctly with the noise even when the bins overlap. In the case of a top-hat window function, we recover the conventional result  \citep{Santos:2015bsa}, assuming a constant system temperature in the band.

There is also a shot noise term in intensity mapping, since the signal requires the existence of galaxies in order to produce the emission lines. However, for HI, this shot noise term is quite small and can be safely neglected \citep{2011ApJ...740L..20G}.

\begin{figure}
\centering
\includegraphics[width=\columnwidth]{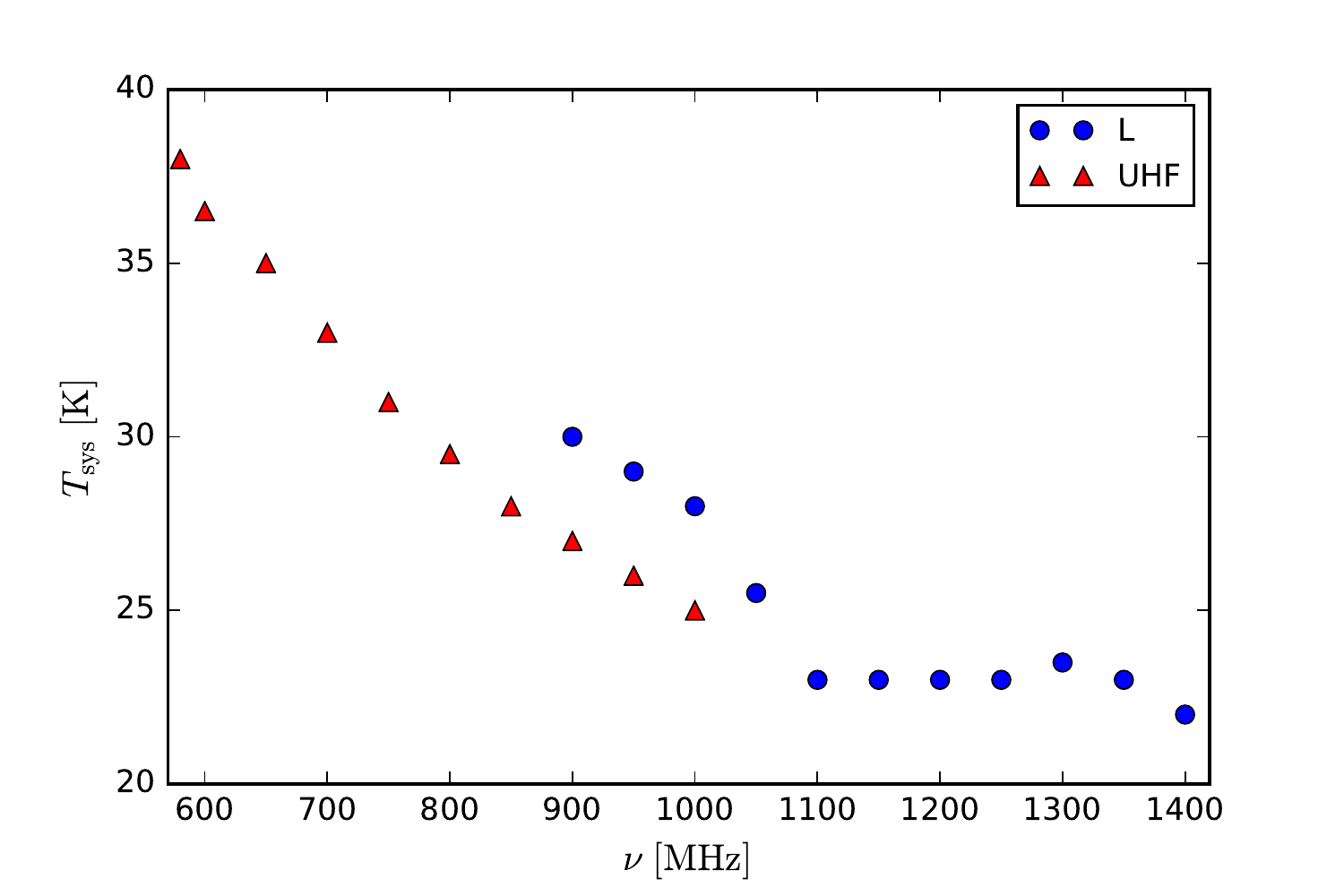}
\caption{MeerKAT system temperature for L-band (blue dots) and UHF-band (red triangles).}\label{fig:MK_t_sys}
\end{figure}

MeerKAT's bands are: 
\bea
\text{L:} &&~~  900<\nu< 1670\,{\rm MHz}, ~~ 0.58>z>0, \\ 
\text{UHF:} &&~~  580<\nu< 1015\,{\rm MHz}, ~~ 1.45>z>0.40.
\eea
Although the total bandwidths are similar, the UHF band will probe a larger physical volume, allowing in principle for better cosmological measurements. 
Figure \ref{fig:MK_t_sys} shows the system temperature for the MeerKAT bands.

\begin{figure*}
\centering
\includegraphics[width=\columnwidth]{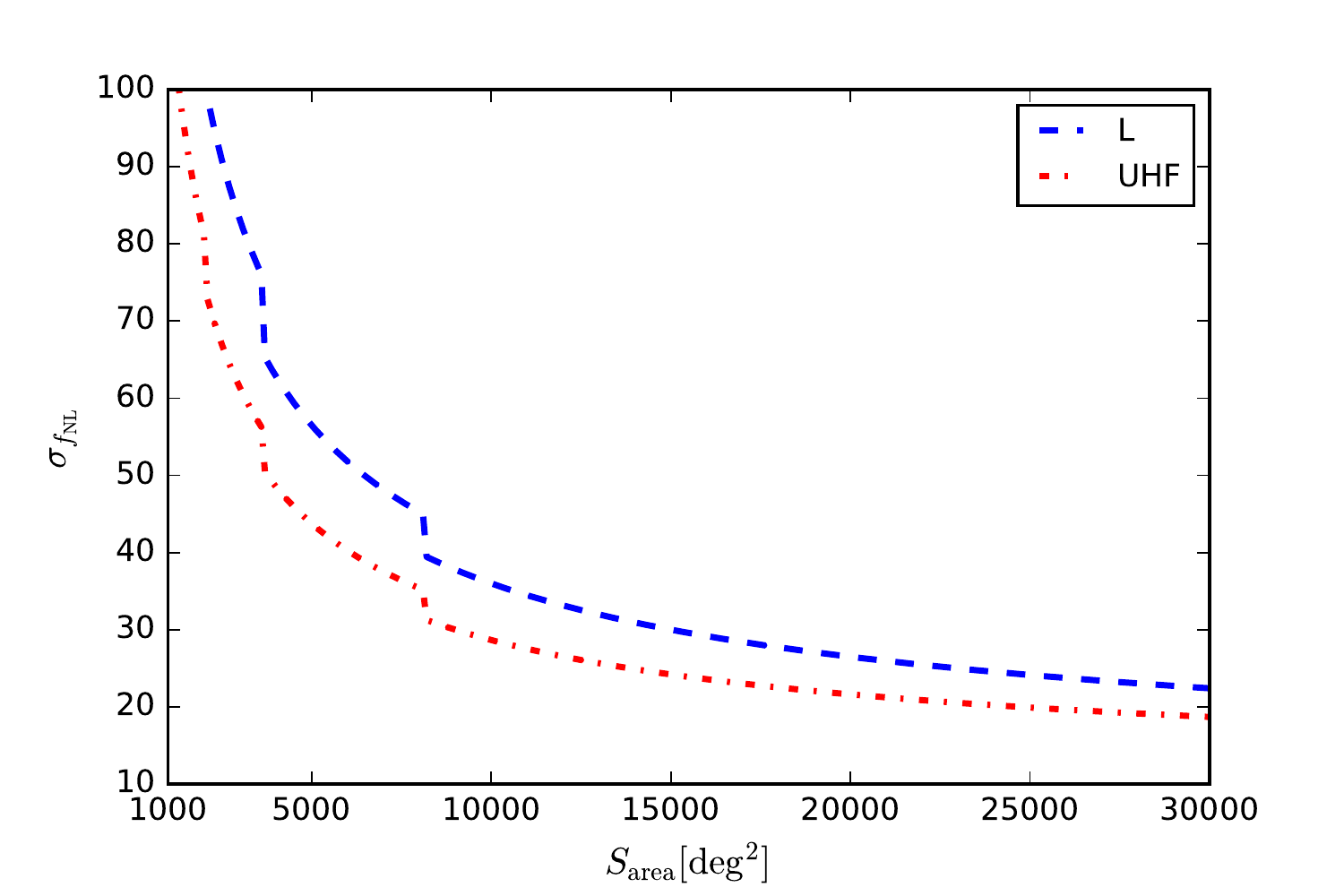}
\includegraphics[width=\columnwidth]{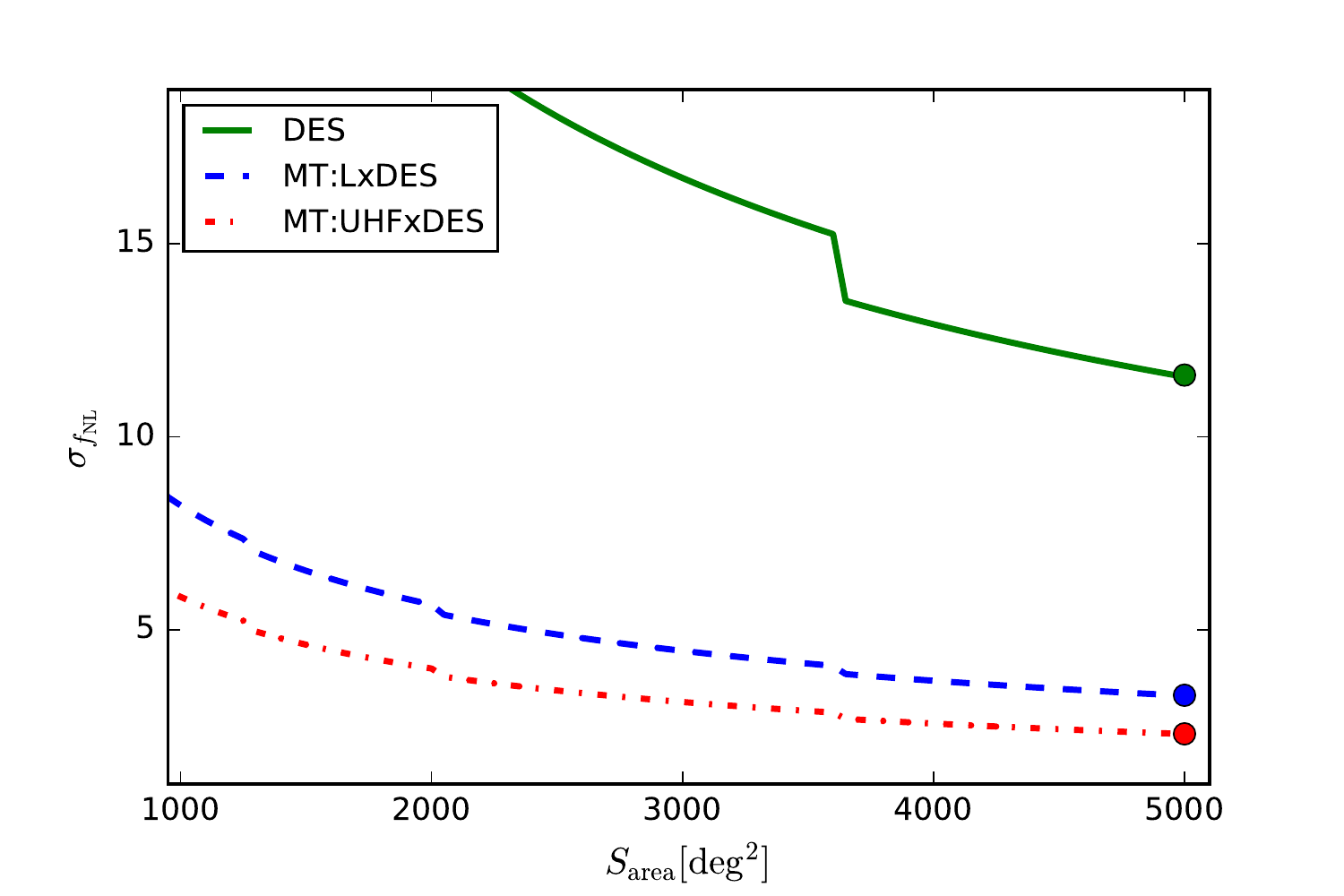}
\caption{Marginal error in measuring $\fnl$, against the surveyed area for different  configurations, with fixed MeerKAT observing time of 4000\,hr. \emph{Left:} MeerKAT with L-band (dashed, blue) and UHF-band (dot-dashed, red). \emph{Right:} DES on its own (solid, green); multi-tracer MeerKAT L-band $\times$ DES (dashed, blue); multi-tracer MeerKAT UHF-band $\times$ DES (dot-dashed, red).}\label{fig:main}
\end{figure*}

\subsection{DES photometric galaxy survey}

DES is a 5000\,deg$^2$ photometric galaxy survey in the southern sky, currently underway. In order to determine the observational details for forecasts, we followed closely the approach taken by  \citet{Alonso:2015uua} (their Section 7.1) for a DES-like photometric galaxy survey. 

Following \citet{Alonso:2015uua}, we adopt a simulation-based model \citep{Weinberg:2002rm} for the Gaussian galaxy bias (see \cite{Clerkin:2014pea} for alternative bias models): 
\bea
 b_{\rm L}^{ G}={1+0.84z}\label{eq:DES-bias}\,.
\eea
Note that this bias model is used only to set the fiducial value in each redshift bin, since we marginalize over bias.

We use an $r'$-band Schechter luminosity function, as given by \citet{Gabasch:2005bb}. We approximate $r'\simeq r$ and use the following parametrisation of the Schechter function: 
\bea
&&\alpha =-1.33\,,~~~~
M_*(z) = -21.49 -1.25 \ln\l1 + z\r, \label{lf1}\\
&&\varphi \l z\r = 2.59 -0.136 z -0.081 z^2\quad \big[10^{-3}\, {\rm Mpc}^{-3}\big].\label{lf2}
\eea  
Here $\alpha$ is the slope of the low end of the luminosity function, and $M_*$ is the magnitude of transition from lower to higher luminosities.
The absolute magnitude $M$ and apparent magnitude $m$ are related by 
\bea
M = m(z)-25-5\log_{10}\frac{ d_{ L}\z}{\rm Mpc} -k\l z\r ,
\eea
where $d_{ L}$ is the luminosity distance and the k-correction is taken  as $k\l z\r\simeq 1.5 z$ \citep{Alonso:2015uua}.

We then use \eqref{lf1} and \eqref{lf2} in the Schechter luminosity function to estimate (with a five-point stencil numerical derivative) the magnification bias \eqref{sg},  which we fit with the polynomial  
\bea
s^{G}(z)&=&0.132+0.259z-0.281z^2\nn\\
&&{}+0.691z^3-0.409z^4+0.152z^5\,.
\eea
We truncated the polynomial once the error between the estimate and the fit to the polynomial was  $<1\%$. 

DES will observe galaxies with magnitude $r<24$ and the redshift distribution of sources that we obtain is modelled as
\bea
\frac{\d n^{G}}{\d z}=22.36 \l\frac{z}{0.57}\r^{1.04} \exp\left[-\l\frac z{0.57}\r^{1.34}\right]\, {\rm arcmin}^{-2}.
\label{desn}
\eea
The overall normalisation agrees with DES Science Verification data \citep{Bonnett:2015pww}. 
Using the relation between $n^G$ and $N^G$, the evolution bias \eqref{be} is given by 
\bea
b_e^{G}=-\frac{\p\ln \left[(1+z)\H\chi^{-2} \d n^{G}/\d z\right]}{\p\ln (1+z)}.
\eea

The noise angular power spectrum for a galaxy survey is dominated by shot-noise \citep{Alonso:2015uua}:
\be \label{eq:noise:gal}
\mathcal N^{G}_{ij}=\frac1{{N^G_i}{N^G_j}}\int\!\d z\,\frac{\d n^G}{\d z}\, {W}\big( z,z_i;\Delta z_i, \s^z_i\big){W}\big( z,z_j;\Delta z_j, \s^z_j\big) \,,
\ee
where $N^G_i$ is the number of galaxies per steradian in the $i$-bin:
\bea
N^G_i=\int\!\d z\,\frac{\d n^G}{\d z}\, {W}\big(z,z_i;\Delta z_i, \s^z_i\big)\,.
\eea
$W$ is the window function centred at $z_i$ with bin size $\Delta z_i$ and photometric redshift scatter $ \s^z_i=\sigma_0(1+z_i)$, with $\sigma_0=0.05$ for DES.
For a photometric survey, the window function is given by a combination of error functions \citep{Ma:2005rc},
\bea \label{eq:Windowg}
W\big( z,z_i;\Delta z_i, \s^z_i\big) =\frac12\left[{\rm erf}\l\frac{z_i+\Delta z_i-z}{\sqrt 2\, \s^z_i}\r-{\rm erf}\l\frac{z_i-\Delta z_i-z}{\sqrt 2\, \s^z_i}\r\right]\!.
\eea
Eq. \ref{eq:noise:gal} is also valid when different redshift bins overlap, which is the case we consider. Note that if we consider a top-hat window function, we recover the result commonly found in the literature $\mathcal N^{G}_{ii}=1/N^G_i$.

\subsection{HI--G cross-noise}

We also take into account the possible shot-noise cross power spectrum.  
This is due to an overlap in the halo mass range which the tracers probe. Even if this is small, it might be important for the multi-tracer, since this is the only component in the noise matrix (\ref{noise:matrix}) corresponding to the cross-correlation between tracers.  We are assuming Poisson noise. Simulations have shown that non-overlapping mass ranges can exhibit off-diagonal shot-noise, and mass-dependent weighting schemes can suppress the total shot-noise contamination \citep{Hamaus:2010im}. However, the error in estimating the cross-correlation between tracers is dominated by the individual noises in each tracer, so that our Poisson assumption is not unreasonable. Then the cross-shot-noise is given by \citep{2015ApJ...812L..22F}
\bea
&&\mathcal N^{ HI,G}_{ij}=\int\!{\d z}\,\bar W\big( z,z_i;\Delta z_i, \s^z_i\big) W\big( z,z_j;\Delta z_j, \s^z_j\big) \frac{T^{ HI}(z)}{\rho_{HI}(z) N^{G}_j} \nn\\
&&~~~\times  \int\!{\d M_h}\,M_{ HI}(M_h)\, \Theta\l M_h,z\r \frac{\d N_h}{\d M_h}(M_h,z)=\mathcal N^{G,HI}_{ji},
\eea
where  $\rho_{ HI}$ is the HI density, $M_{ HI}$ is the mass of HI in a halo of mass $M_h$, and $\d N_h/\d M_h$ is the halo mass function. If the halo masses probed by the two surveys overlap, then $\Theta(M_h)=1$, otherwise it is zero. For further details on the halo mass range for HI intensity mapping, see \cite{Santos:2015bsa}. The mass range for a photometric survey is found by matching the number of galaxies given by the halo mass function with the number given by the selection function. 
Note that the two windows have similar shapes but different normalizations, i.e., $\bar W\propto W$. While $W$ is given by Eq. (\ref{eq:Windowg}), $\bar W$ is the same as in equation (\ref{eq:N_HI}) and is normalised to 1. 

Including all noise contributions, we can write the multi-tracer noise angular power spectrum matrix as
\bea \label{noise:matrix}
{ \cal N}^{AB}_{ij}
=
\begin{pmatrix}
   {  \cal N}^{HI}_{ij} & { \cal N}^{HI,G}_{ij} \\ & & \\
    { {\cal N}^{G,HI}_{ij}} & { \cal N}^G_{ij} 
\end{pmatrix}\,.
\eea
Note that it is independent of the multipole $\ell$.

\begin{figure*}
\centering
\includegraphics[width=\columnwidth]{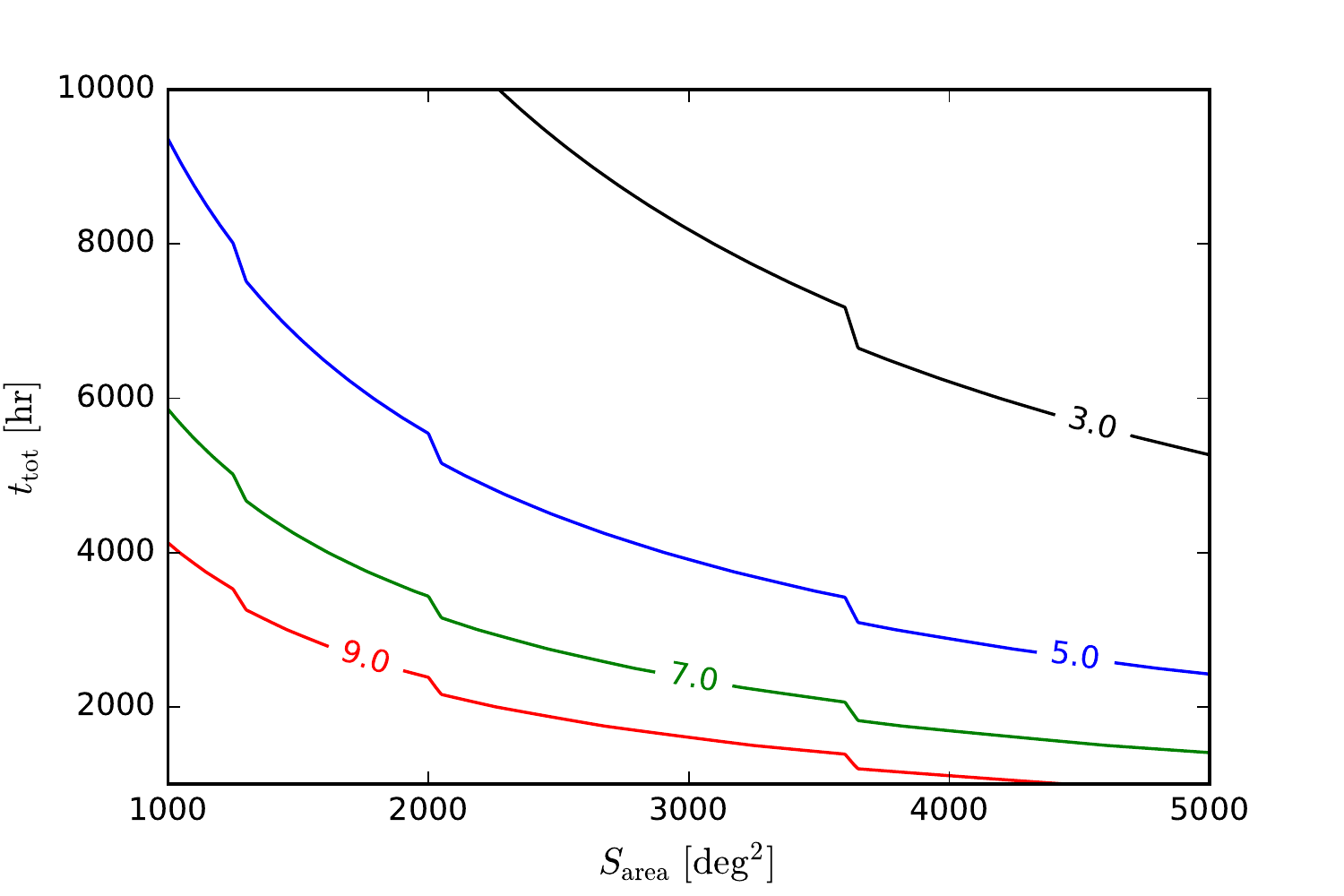}
\includegraphics[width=\columnwidth]{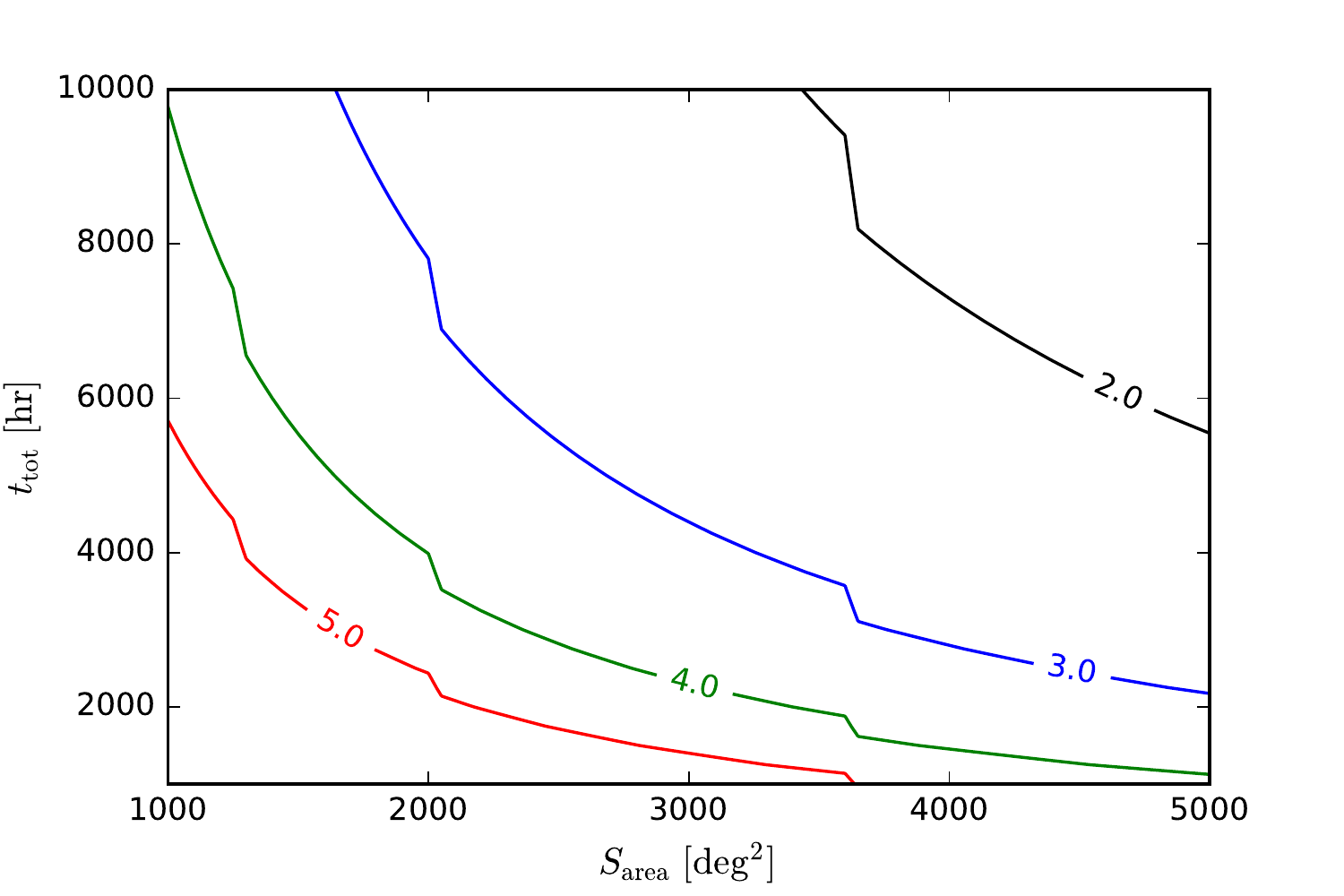}
\caption{Contour plot of $\s\l\fnl\r$ against survey area and observing time, using the multi-tracer with MeerKAT and DES. {\em Left:} L-band $\times$ DES. {\em Right:} UHF-band $\times$  DES. }\label{fig:cs:MKL_DES}
\end{figure*}

\section{Results}\label{sec:results}

We perform the Fisher forecast analysis as described in section \ref{sec:fisher} for the set of parameters
\bea \label{para}
  \vartheta_\alpha=\{\ln A_s,\ln\odm,\fnl,\ln n_s,\ln \ob, w, b^A_i,   \varepsilon_{\rm WL}, \varepsilon_{\rm GR} \}, 
\eea  
where $\odm$ is the density parameter of cold dark matter, $\ob$ is the density parameter of baryonic matter and $w$ is the dark energy equation of state parameter. We assume a fiducial concordance flat cosmology with $H_0=67.74\,$ km/s/Mpc, $\odm=0.26$, $\Omega_{b}=0.05$, $A_s=2.142\times 10^{-9}$, $n_s=0.967$, $w=-1$ and $\fnl=0$. The bias parameters $b^A_i$ in each bin have fiducial values shown in Fig. \ref{fig:bias}. The last two parameters in \eqref{para} have fiducial values $\varepsilon_{\rm WL}=1= \varepsilon_{\rm GR}$, and are defined so as to isolate the weak lensing and general relativistic terms in \eqref{eq:angGS}:
\bea
\Delta_\ell^A= \Delta_\ell^A({\rm density+RSD})+ \varepsilon_{\rm WL}\Delta_\ell^A({\rm WL})+ \varepsilon_{\rm GR} \Delta_\ell^A({\rm GR}).
\eea
These parameters take into account that we do not have full knowledge of the evolution and magnification biases in \eqref{eq:angGS}. 

To compute the multi-tracer angular power spectrum we modified the  code \cs \, \citep{Challinor:2011bk} so that it computes both auto- and cross-tracer correlations with the correct selection function. We also changed it to compute the correct evolution bias of each tracer and to have different window functions as options. The output is in the same format as \cs.  The modified code is available on GitHub\footnote{https://github.com/ZeFon/CAMB\_sources\_MT}.

We computed forecasts for the single surveys and the combined surveys, with the following configurations:

\paragraph*{MeerKAT L-Band:}
24 bins of  width 20\,MHz between 1380\,MHz and 920\,MHz; sky coverage from 1000 to 30000\,deg$^2$; a smooth top-hat window function.

\paragraph*{MeerKAT UHF-Band:}

21 bins of width 20\,MHz between 1000\,MHz and 600\,MHz; sky coverage from 1000 to 30000\,deg$^2$; a smooth top-hat window function.

\paragraph*{DES:}

{8 bins in the redshift range $z=0-2$, each with the same number of galaxies; sky coverage from 1000 to 5000\,deg$^2$; an error window function.

\paragraph*{Multi-tracer: L-Band {$\times$} DES:}

4 bins which coincide with the first 4 bins taken for DES alone; 
sky overlap of 1000 to 5000\,deg$^2$; an error window function.

\paragraph*{Multi-tracer: UHF-Band {$\times$} DES:}

5 bins between $z=0.40$  and 1.45;  
sky overlap of 1000 to 5000\,deg$^2$; an error window function.
\\

The minimum $\ell$ used in our forecasts depends on the surveyed area: $\ell_{\rm min}= 1+$ the integer part of  $\pi/\sqrt{S_{\rm area}}$. For the maximum $\ell$, we only consider information in the Fisher matrix if the scales  are within the linear regime, as defined by  \citep{Smith:2002dz}
\be
k_{\rm NL}(z)=k_{\rm NL,0}\l1+z\r^{2/(2+n_s)}, \quad\mbox{with}\quad k_{\rm NL,0}\simeq 0.2 h\,{\rm Mpc}^{-1} .
\ee
Using the Limber approximation, $\ell_{\rm max}\simeq \chi k_{\rm NL} $. Each redshift bin has its own corresponding $\ell_{{\rm max}, i}$. We therefore neglect the information coming from the $i$-th bin in the sum of the Fisher matrix  (\ref{eq:fishercl}) when $\ell>\ell_{{\rm max}, i}$. This will only be necessary for the low redshift bins. For higher redshifts, we impose the global maximum $\ell_{{\rm max}}=300$, since the additional information from higher $\ell$ (within the linear regime), provides very little improvement on the constraints.

The main results of this paper can be seen in Fig. \ref{fig:main} and Table \ref{tab:table}, where we fix the MeerKAT observational time at 4000 hours. The error on $\fnl$ has been marginalised over the other parameters in \eqref{para}. None of the surveys on their own can match the accuracy on $\fnl$ of Planck (although DES is close).  
But with the multi-tracer technique, MeerKAT combined with DES over an overlap area of $\sim 4000- 5000\,$deg$^2$, improves significantly on the Planck  $\sigma(\fnl)$ -- for both bands.  
(Note that the steps seen in the curves in Fig. \ref{fig:main} come from the fact that as the surveyed area decreases, the minimum accessible $\ell$ increases.)

\section{Conclusions}\label{sec:conc}

\begin{table}
\caption{Marginal errors on $\fnl$ for HI intensity map surveys with MeerKAT L- and UHF-bands, a DES photometric survey, multi-tracer analyses combining DES and each MeerKAT band, and the two multi-tracer analyses combined. (We assume 5000\,deg$^2$ survey area, and 4000\,hr MeerKAT time.)}
\centering
\begin{tabular}{l c}
\hline
&$\s(\fnl)$\\
\hline &\\
MeerKAT L-Band  & 56.5\\
MeerKAT UHF-Band   & 43.8\\
DES  & 11.9\\
MT: MeerKAT L-Band $\times$ DES & 3.6\\
MT: MeerKAT UHF-Band $\times$ DES & 2.3\\
\hline
\end{tabular}
\label{tab:table}
\end{table}

Table \ref{tab:table} summarises the marginal errors on $\fnl$ for the individual and multi-tracer cases, using a survey area of 5000\,deg$^2$ and MeerKAT's integration time of 4000\,hr.
DES with the MeerKAT L-band is two times better than Planck, while DES combined with the UHF-band  improves on Planck by a factor of three. In both cases, the multi-tracer is forecast to beat Planck on $\fnl$ within the next few years.

The right panel of Fig. \ref{fig:main} {also} shows that the multi-tracer technique is powerful enough to improve the forecast error on $\fnl$ even with a smaller surveyed area, as can be seen by comparing DES on its own to the multi-tracer of DES with MeerKAT. 
This can be contrasted with single-tracer measurements, which require larger volumes to reduce the error bars. 

Figure \ref{fig:cs:MKL_DES} shows how our results vary with integration time and surveyed area. Contours are plotted for $\s\l\fnl\r$ when both the overlap survey area and the MeerKAT integration time are varied.  Even a survey overlap area of $\gtrsim 2000\,$deg$^2$ and a MeerKAT observation time $\gtrsim 2000\,$hr suffices to give an improvement over Planck-level accuracy with the UHF-band $\times$ DES (right panel). The same area and integration time with the L-band $\times$ DES (left panel) gives an improvement on full DES (with 5000\,deg$^2$). 

We have assumed that  all multipoles down to $\ell=3$ can be used when considering a 5000\,deg$^2$ survey. If the largest scales are not accessible the result worsens as we can see in Fig. \ref{fig:sigfnl:ellmin} (left). The effect is more prominent for the single tracer case, as shown for DES. In the multi-tracer case, the accuracy is only mildly degraded.

\begin{figure*}
\centering
\includegraphics[width=\columnwidth]{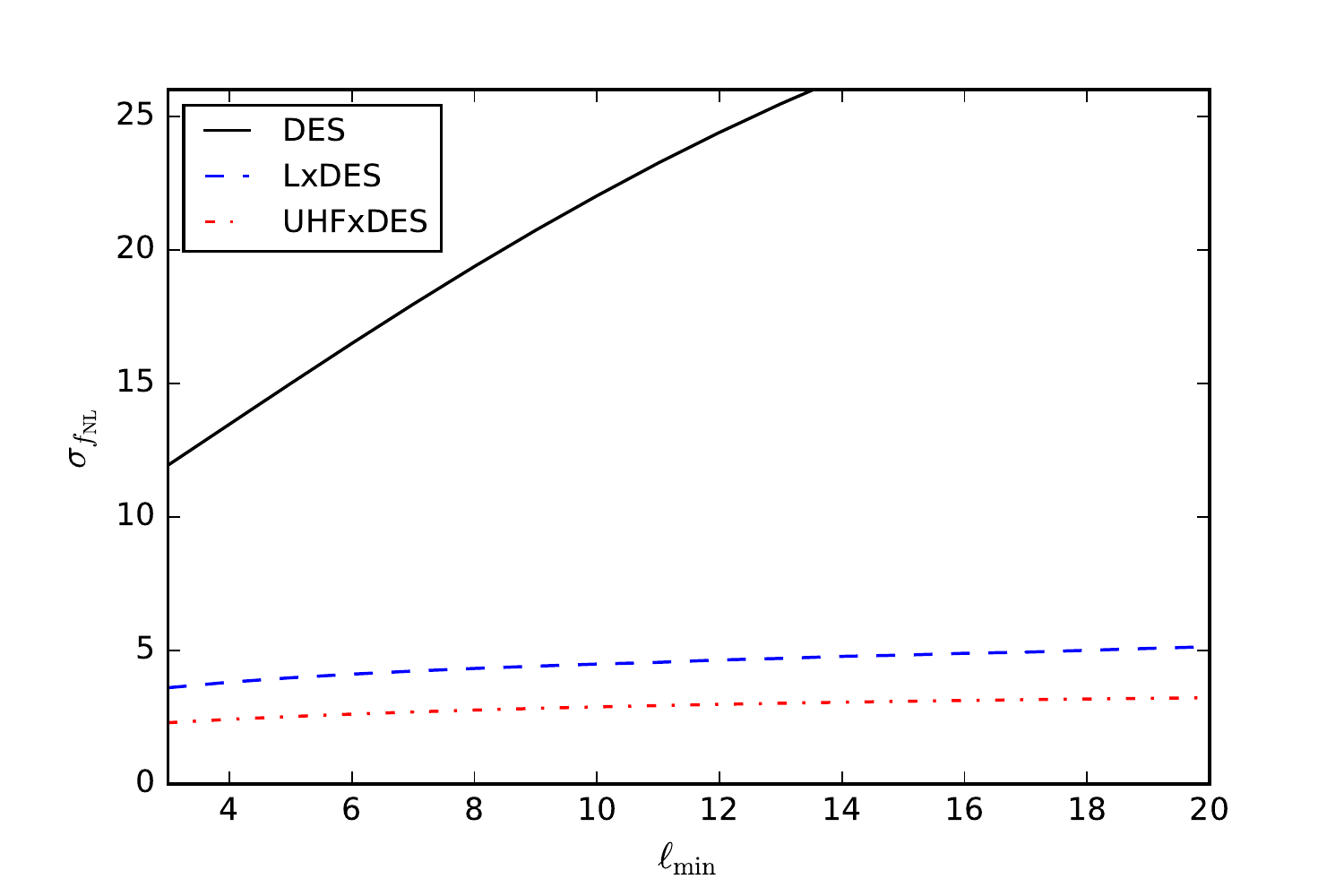}
\includegraphics[width=\columnwidth]{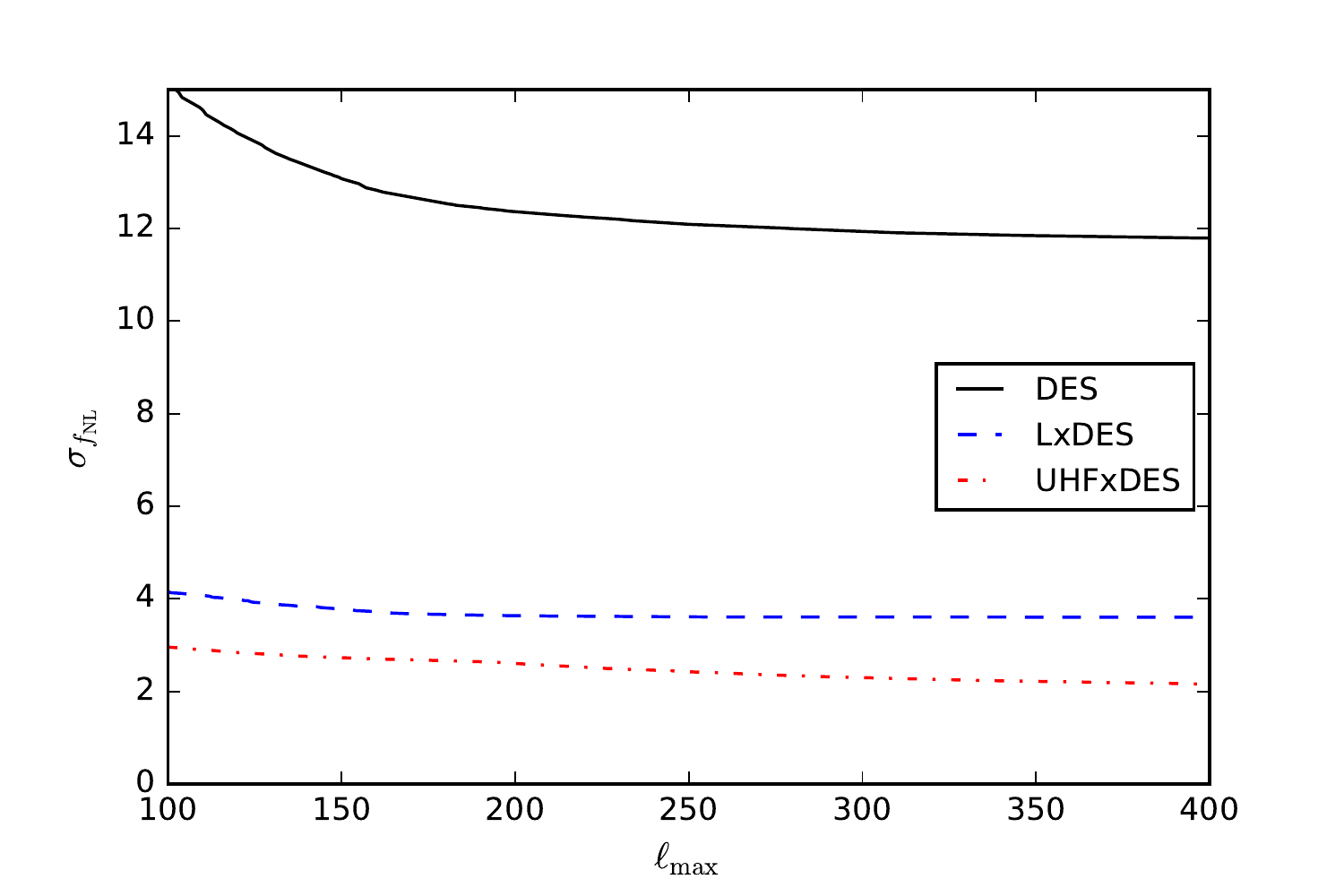}
\caption{$\s\l\fnl\r$ as a function of the minimum multipole ({\em left}) and maximum multipole ({\em right}) used in the Fisher forecast (with 5000\,deg$^2$ survey area and 4000\,hr MeerKAT time).}\label{fig:sigfnl:ellmin}
\end{figure*}

We chose to truncate the sum in the Fisher matrix at $\ell=300$, even when higher $\ell$ would still correspond to linear scales. Although this choice may seem arbitrary, we can see in Fig. \ref{fig:sigfnl:ellmin} (right) that not much more information is added for $\ell\gtrsim150$.

We use models for the HI and galaxy bias to provide the fiducial values in each redshift bin. The uncertainties in bias modelling can be mitigated by marginalising over the bias in each redshift bin.  
The results of \cite{Alonso:2015sfa,2015ApJ...812L..22F} indicate that uncertainties in the bias are less important than those in the magnification bias and evolution bias. We also incorporate uncertainties in these by marginalising over the parameters  $\varepsilon_{\rm WL}$ and $\varepsilon_{\rm GR}$ respectively. 

We do not include foregrounds and observational systematics in creating the maps in the full covariance matrix -- but the multi-tracer technique includes cross-correlations and thereby lessens the impact of individual systematics and of foreground residuals. 

We conclude that the best contemporary radio and optical surveys, i.e., MeerKAT and DES, when combined via the multi-tracer technique, can improve on the Planck error bars for $\fnl$, well before the next-generation surveys deliver data.
This is important not only for improving on Planck -- but also because it can serve as a ``proof of concept" for the multi-tracer technique applied to primordial non-Gaussianity. The MeerKAT--DES multi-tracer will effectively be a pathfinder for radio--optical multi-tracing with next generation surveys, such as SKA--LSST or SKA--Euclid.

~\\ \noindent{\bf Acknowledgements:}
The authors are supported by the South African Square Kilometre Array Project and National Research Foundation. RM is also supported by the UK Science \&\ Technology Facilities Council, Grant No. ST/N000668/1.

\bibliographystyle{mn2e}

\label{lastpage}

\end{document}